# Cyber-Human System for Remote Collaborators

Srikanth Jonnada, Ram Dantu, Ishan Ranasinghe, Logan Widick, Mark Thompson and Janice A. Hauge

University of North Texas, Denton, TX, USA (E-mail: ram.dantu@unt.edu)

**Abstract.** With the increasing ubiquity of technology in our daily lives, the complexity of our environment and the mechanisms required to function have also increased exponentially. Failure of any of the mechanical and digital devices that we rely on can be extremely disruptive. At times, the presence of an expert is needed to analyze, troubleshoot, and fix the problem. The increased demand and rapidly evolving mechanisms have led to an insufficient amount of skilled workers, thus resulting in long waiting times for consumers, and correspondingly high prices for expert services. We assert that performing a repair task with the guidance of experts from any geographical location provides an appropriate solution to the growing demand for handyman skills. This paper proposes an innovative mechanism for two geographically separated people to collaborate on a physical task. It also offers novel methods to analyze the efficiency of a collaboration system and a collaboration protocol through complexity indices. Using the innovative Collaborative Appliance for Remote-help (CARE) and with the support of a remote expert, fifty-nine subjects with minimal or no prior mechanical knowledge were able to elevate a car for replacing a tire; in a second experiment, thirty subjects with minimal or no prior plumbing knowledge were able to change the cartridge of a faucet. In both cases, average times were close to standard average repair times, and more importantly, both tasks were completed with total accuracy. Our experiments and results show that one can use the developed mechanism and methods for expanding the protocols for a variety of home, vehicle, and appliance repairs and installations.

## 1 Introduction

Many consumers lack either the technical skills to complete their own repair, or don't wish to struggle completing their own repair; one needs cautionary and interactive technical assistance that online videos or documentation may fail to provide. Many times a quick, simple fix can prevent further damage, but these quick fixes also need technical expertise that is not available unless an expert is physically present, especially if something goes wrong as they often do, and such physical presence generally carries a high price [1][2][3].

It is important to consider ways to support the increasing demand for expert repair services. Transportation to and from repair sites generally remains time-consuming and can be expensive; working remotely has thus become a popular choice for those relying on digital technologies, whereby a remote expert can assist with digital tasks from afar. Performing a task with the guidance of experts from any geographical location provides an apt solution to the growing demand for the skilled workforce. This paper proposes an innovative mechanism for two geographically separated people to collaborate on a physical task. It also offers a novel method to measure the complexity of a human collaboration protocol that helps in designing efficient remote collaboration systems and human protocols.

## 2 Motivation

The primary motivation for this work is the lack of efficient mechanisms for geographically separated individuals to collaborate on a physical task and the lack of methodologies to measure and minimize the complexity of such a collaboration. Human communication is crucial for many systems; miscommunication or misunderstanding of a message can cause unexpected delays and costly damages. A protocol is a set of rules or a process followed while performing an activity that facilitates accurate communication and successful operation of the systems [4][5]. For instance, in 911 emergency dispatch systems [6] and air traffic control systems [7] participants follow a defined protocol and exchange crucial information to gain knowledge about the situation and establish a common understanding. Designing a straightforward human collaboration protocol can facilitate an accurate and easy exchange of information among individuals.

## 3 Background on Human Collaboration and Existing Systems

Grounding is a concept that comprises the collection of mutual knowledge, beliefs, and assumptions deemed to be essential for communication among individuals [8]. Individuals can establish grounding with words and with actions; a positive action can act as an implicit acknowledgment [9]. Situational awareness is a process where people obtain complete knowledge about the work environment and tasks being executed; this helps people understand the situation and make the right decisions at the right time [10]. An expert helps a novice execute a task; throughout this document, we refer to the expert as a helper and the novice as a worker. When the helper and worker are physically together, situational awareness is straightforward: the helper can gain knowledge of the worker's actions and task environment by visual inspection. They can establish grounding through voice communication and gestures. However, situational awareness and grounding can be a challenge when individuals are geographically separated. In a remote collaboration, voice communication, visual information [8][11][14][15], and gestures [12] are crucial for situational awareness [10] and the medium of communication [8] impacts the grounding to a greater extent.

## 3.1 Grounding

For a team to work together and successfully execute a task, the members should share knowledge and information. This requires that every member of the team maintain a common understanding throughout the task they are performing.

As per Clark et al. [8], a team of two people working together on a task must establish a common understanding of both content and process. For this to happen, they must share a vast amount of information and update their common ground continuously. The speaker must be sure that the listener understands him; for this, he should look for positive evidence of understanding from the listener. Acknowledgments and initiation of the next step in a process are forms of positive evidence that form a base for common grounding.

As Clark et al. [8] state, it is necessary for the participants in a collaboration to come to a common ground on the identity of the objects as quickly as possible; there are various techniques available for this. The participants can provide definite or indefinite descriptions of the objects and can use gestures and referential installments. Choosing the right medium of communication can help reduce the collaboration effort.

## 3.2 Situational Awareness

When a helper is guiding a worker in accomplishing a task, the helper should be aware of the work environment and every action of the worker. This awareness provides information to the helper and allows him or her to guide the worker efficiently and effectively. Situational awareness (SA) plays a crucial role in human decision making. SA is widely used by major industries today; for example, in the aviation industry, SA is critical in the safe operation of planes. The air traffic controller must have an awareness of multiple factors relating to the route, position, external conditions, and aircraft-related details to safely instruct pilots.

Endsley [10] states that SA is critical for human dynamic decision making in operating the crucial and complex systems that technology has brought us. Even everyday activities require situational awareness of tasks. In many task environments, the workers must work in a dynamically changing situation and make decisions in a narrow period of time. Understanding a situation through SA is vital before proceeding with actions. The process of SA allows recognition of the appropriate cues and sets expectations for future states of the situation. Awareness of critical information in a dynamically changing environment is crucial for a worker to act on a task. The degree to which a system provides a worker with information about the situation impacts the worker's ability to achieve SA. Endsley [10] hypothesizes a set of features of a system that could positively impact the SA:

the representation of goal-oriented information, projection of critical cues, the capability to process information in parallel, avoidance of unnecessary information, and reduced presentation of non-critical cues.

A remote-helper lacks the same SA of a bystander. It is a challenge to design a system that mediates a remote-helper and worker and provides the required SA. The way the system presents the acquired information to the helper is also crucial and can impact the SA of the helper.

In sum, common grounding and SA are essential for successful collaboration between a remote-helper and a worker. The capabilities of the communication system can influence the quality of common grounding and SA. We discuss these capabilities in the next section.

## 3.3 Collaboration System Capabilities that Impact Common Grounding and Situational Awareness

### 3.3.1 Visual Information

Kraut et al. [13] performed a study to understand the influence of visual information as a resource for conversation in a collaborative physical task. The study utilized a head-mounted camera and a microphone as sources of visual and audio information while participants undertook a bicycle repair task. They observed that side-by-side workers have better performance than a participant and a remote-helper as the subjects can refer to objects more quickly with deictic expressions in a side-by-side condition. The head-mounted camera of the worker only provided a partial view of the task environment and the remote-helper had no control over the view. Kraut et al.'s work illustrates that visual information is valuable in remote collaboration for maintaining SA and establishing grounding.

A study conducted by Gergle et al. [14] shows that the benefits of visual information on building common grounding and SA depend on the technologies used and their characteristics. The study performed experiments to analyze the impact of visual features that include delay, field of view, perspective and view control on conversational grounding, and SA. They found that visual feedback delayed more than 950ms can disrupt coordination. Moreover, they discovered that when two collaborating parties have the same spatial perspectives, they can perform the task more quickly than when they have unaligned perspectives. With unaligned perspectives, users are not able to use spatial descriptors and have to use more verbally-vague descriptions. In addition, the authors show how the scale of area of the view impacts the SA; with more area, users have a difficult time tracking the states of the task, thereby increasing the number of utterances.

Fussell et al. [15] performed a study to understand the impact of different sources of visual information. They include heads and faces, bodies and actions of workers, task objects, and the work environment of the remote collaboration on a physical task. The study tested collaboration in assembling a robot in five different media conditions: side-by-side, head-camera with eye-tracking, scene camera, scene camera plus head camera, and audio-only. The results indicate that side-by-side is faster than all other media sources, followed by a collaboration with a scene camera. The authors also found that scene-oriented cameras can provide significant information compared to audio-only communication and the head-mounted camera. The scene-oriented camera provides a view of the worker's environment and provides visibility of his actions. Even though this may not provide close-up views of the worker's actions, the scene-oriented camera provides better SA than the head-mounted camera. A greater quality of visual information provides greater SA.

Ranjan et al. [16] conducted a study to analyze the impact of controlled visual information on the efficiency of collaboration on a physical task. The study performed experiments using three camera-controlled conditions, which included helper-controlled, operator-controlled, and automatic-controlled cameras. The authors found higher effectiveness in collaboration with helper-controlled cameras compared to the other two conditions.

Clearly, visual information can significantly improve the common grounding and situational awareness between the remote-helper and worker. With either scene-oriented or head-mounted cameras the helper cannot gain information about the complete work environment that is vital to collaboration. The remote-helper needs complete control over the camera to fetch the desired visual information. A system with remotely controlled mobility, panning, tilting, and zooming of the camera provides required visual information to the remote-helper and helps attain the SA, ultimately benefiting the collaboration.

### 3.3.2 Gestures

During a conversation or collaboration, along with oral information, participants provide gestures with hand or facial expressions to convey information to the other party. Gestures may provide a faster and easier way of communication than words alone might provide. In a remote collaboration, gestures can help establish common grounding between the helper and worker.

Alem and Li [17] performed a study to understand the relative effect of overlaying-hands and cursor-pointer representation on the task performance. The study found similar effectiveness of both gesture representations, but the participants preferred cursor pointer to hand gestures. Ou et al. [18] implemented gestures over a video environment, where a remote-helper draws over the video feed he receives and the combined video feed is transmitted back and displayed on the screen in the worker's environment. A study by Fussell et al. [19] analyzed drawing-over-video and cursor-over-video systems; the authors claim that drawing-over-video will lead to communication identical to that of side-by-side collaborations. The gestures over video streams require a full-duplex video that increases bandwidth consumption and is subject to latencies that can disrupt the collaboration. Yamazaki et al. [20] have developed a remotely controlled laser pointer for remote gestures.

The collaboration systems should not be constrained to a fixed environment; one should be able to use them for any task and environment. Gestures that can point precisely to the target object and provide a clear indication to a worker will help the collaboration. Laser pointers can serve as precise gestures that can be focused accurately even on small target objects.

## 3.4 Existing Collaborative Systems

**Table 1: Existing collaborative system**

| Existing Collaborative System | Proposed System |
|---|---|
| Huang et al. [21]<br><br>Both the worker and helper wear a helmet with a camera and a near-eye display. This system implements gestures over a video feed, where the helper draws hand gestures in the air that are captured by the head-mounted camera and overlaid on the video feed that is transmitted back to the worker. | The head-mounted cameras do not provide significant benefit on SA and studies [15] have shown that they are not more efficient than audio-only devices. In the proposed system, the helper can remotely control the pan and tilt movements of the camera to obtain visual information of the task environment. |
| Ou et al. [18]<br><br>Develops a DOVE (Drawing Over Video Environment) technique in which the video feed displays the worker's work area on the tablet of the remote-helper; the helper draws on the tablet that is overlaid and displayed on the worker's desktop PC. | The DOVE system is constrained to a desk environment and is not practical when a task involves outdoor activities. The proposed system is not constrained to a desk environment. In order to test the proposed system some of the experiments were conducted outside. |
| Kuzuoka et al. [22]<br><br>Develops a device that consists of a robot base, notebook PC, WLAN unit, wireless video transmitter, and laser with actuators. Video and voice are transmitted using a satellite communication system that is an inter-university network. | Kuzuoka et al.'s system had limitations with its degree of freedom for both visual information and gestures and field of view for visual information. This caused collaboration issues, whereby the instructor had to exert additional effort relocating the robot completely, thus adding to the collaboration delays. Also, the authors utilized satellite communication for the collaboration, which caused |

| | transmission delays, and this greatly impacted the performance of the collaboration. The proposed system overcomes this limitation by integrating pan and tilt movements to the camera unit and the gesture unit. |
|---|---|
| Robert et al. [24]<br><br>Develops a mechanism where a remote-helper uses a tablet and the worker uses a head-mounted camera and near-the-eye display for collaboration. This system uses video-overlay for gestures. | Robert et al.'s approach has the same difficulties with SA as other head-mounted cameras used for visual information, which prove to be no more efficient than audio-only devices. By providing control over pan and tilt movements of the camera to the helper, in the proposed system the helper does not need to provide the instructions to move the camera to obtain visual information of the task environment. |

# 4 Infrastructure

To facilitate the helper with all the needed capabilities, grounding, SA, and gestures, we developed a novel collaboration system called Collaborative Appliance for Remote-help (CARE).

## 4.1 Architecture of CARE

CARE consists of multiple components, with a Raspberry Pi board serving the computing requirements. Software components on the Pi take a user's input and communicate with different hardware modules. The architecture diagram in Figure 1 provides a snapshot of various layers of elements in the system.

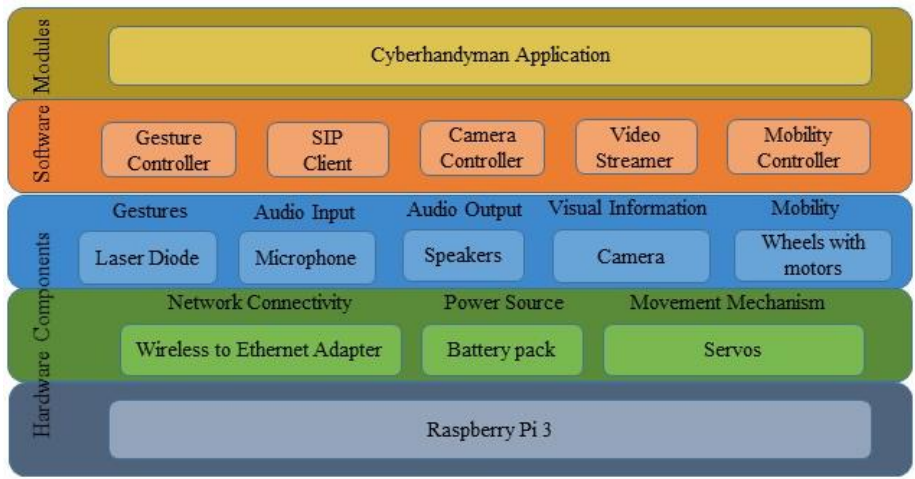

**Figure 1: Layered architecture of CARE system showing all the software and hardware components of the device**

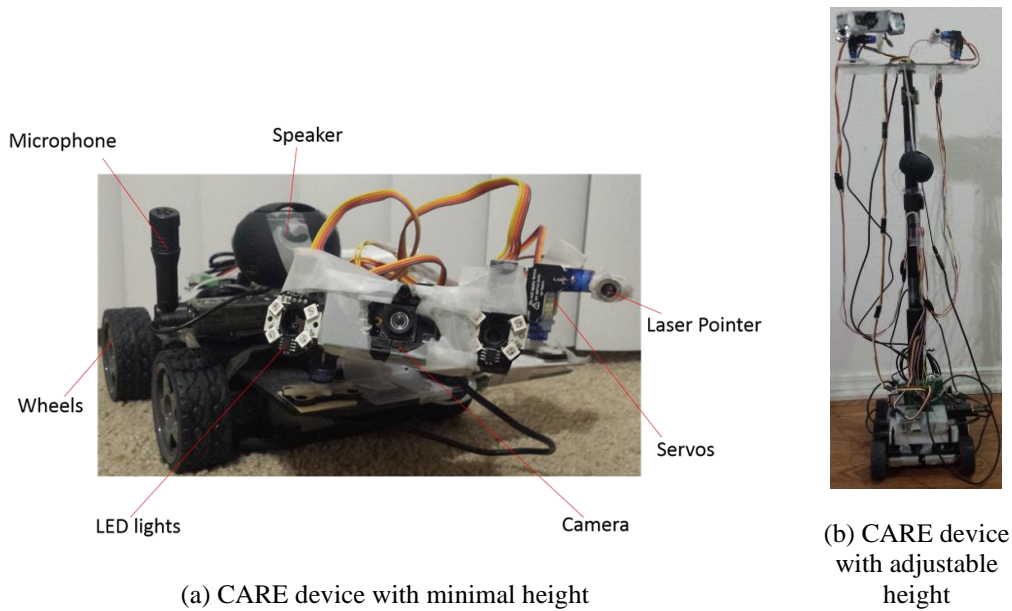

(a) CARE device with minimal height

(b) CARE device with adjustable height

**Figure 2: CARE devices**

## 4.2 Building the Capabilities

### 4.2.1 Visual Information

Visual information acts as an excellent source of information to build the required SA. This information can provide implicit feedback to the helper; it conveys to the helper the extent to which the worker understood his instructions and provides the status of the worker's action. CARE utilizes a high-definition web camera and a video-streaming application to transmit video to a helper's remote location. It provides view control capabilities through the helper's dashboard. The helper can remotely control the pan and tilt movements of the camera to obtain visual information of the task environment. It also provides the zoom in and out feature for a close-up view of objects and manual adjustment of camera focus. For overcoming low light issues, the instrument is equipped with LED lights that the remote-helper can turn on and off and control light intensity.

### 4.2.2 Voice Communication for Common Grounding

The remote-helper and the worker must communicate during the collaboration to establish common ground. This grounding requires a two-way audio channel. The instrument equips a Voice over Internet Protocol (VoIP) client that registers to a server called IP Private Branch Exchange (IP-PBX). The helper also uses a VoIP client registering the same IP-PBX that the instrument registers. This communication system uses Session Initiation Protocol (SIP) for establishing voice sessions and Real-time Transfer Protocol (RTP) for transmitting media. This communication system uses Asterisk-based FreePBX as the SIP registrar and an open-source SIP application, Linphone, as both helper and worker clients. For audio input and output, CARE uses a speaker and a microphone connected to its USB ports. Figure 3 shows the topology of the voice communication system used for the instrument.

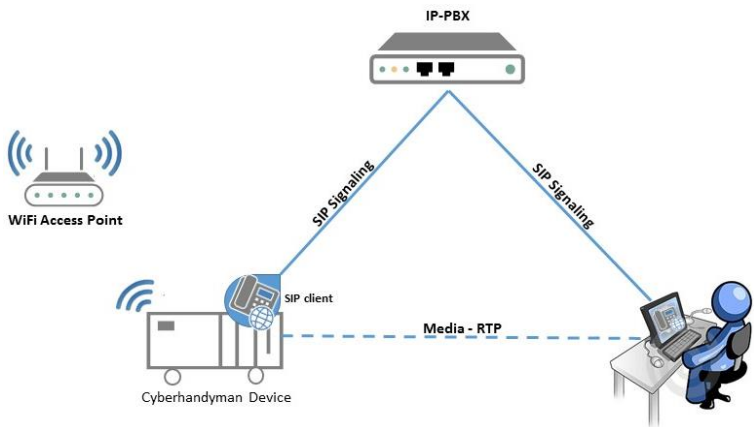

**Figure 3: Voice communication topology, with a SIP client and call processing server.**

### 4.2.3 Remote Gestures

Gestures help the collaborators with grounding. Providing remote gestures is a challenge; the gestures should be precise and should be able to point at small targets. A laser pointer can fulfill these requirements; therefore, we mount a laser diode over a set of servos to enable horizontal and vertical movements. The remote-helper can control the pointer and provide gestures to the worker.

### 4.2.4 Mobility

The mobility of our instrument helps the remote-helper to gain more information about the worker's environment. All the components are stationed on a wheeled platform, with wheel motors connected to a motor controller. Using the software modules and user-interface, the remote-helper can control the movements of the device and can attain or provide information.

### 4.2.5 Wireless Network Connectivity

For the helper to fetch visual information, establish voice communication, provide gestures, and control the movements of the instrument, the helper requires a communication link. Our instrument utilizes a wireless router to connect to a network. The router has pre-configured connections and connects all the components in the instrument. A battery pack on the instrument provides power to the router. In this work, we used a closed wireless LAN network, connecting the instrument, remote-helper, and the IP-PBX.

### 4.2.6 Software Components

CARE consists of a controller application and provides a web interface for the helper, as shown in Figure 4. The controller application directs all the incoming requests to the appropriate component of the device. The controller displays a video feed to the helper and provides all controls to operate resources in the instrument. The user can provide keyboard and mouse inputs to control movements of the camera and laser pointer. Moreover, the web page provides control options to adjust light intensity, adjust the camera focus, turn the laser on or off, control the laser through the mouse, zoom in and out, capture pictures from the camera, and bring the laser to the field of view by the click of a button. The right frame provides controls to adjust the speed of the device and bring the device to an immediate stop; the users can control movements of the device using arrows keys on the keyboard. When the helper clicks on the video frame, he can use arrow keys on the keyboard to control the pan-tilt movements of the camera; when the helper clicks on the right frame of the page, he can use arrow keys to control wheels of the device.

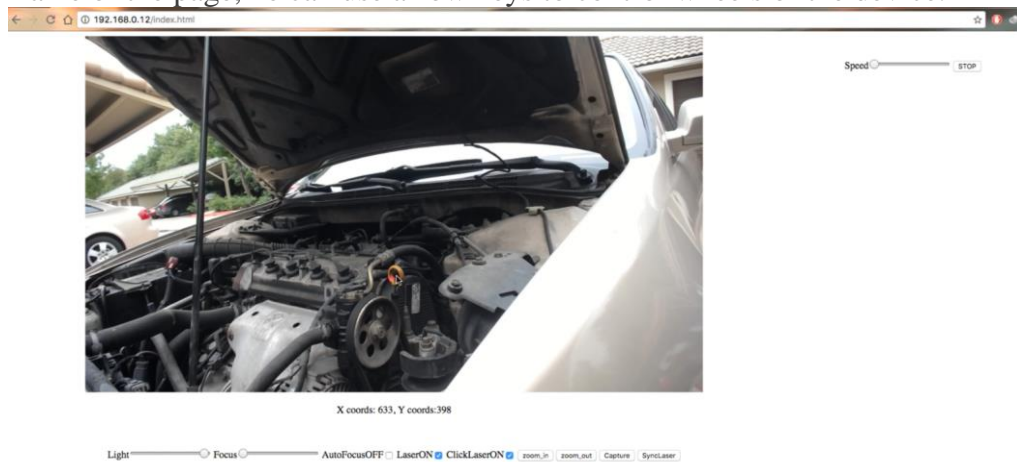

**Figure 4: Display of work environment at the helper end (*Web Interface of the controller application*)**

# 5 Protocol Complexity Theory

If the outcome of a task is positive and meets expectations, the task is said to be successful, and if the outcome is negative, the task is said to be a failure. However, defining the success of a helper-worker collaboration is not based solely on the outcome; the efficiency of the collaboration should be considered as well. When a helper and a worker collaborate to complete a task, efficiency requires that they follow a protocol though an exchange of requests and responses. The success of the collaboration depends on the efficiency of this protocol. When a helper can establish common ground with the worker with minimal complexity, the protocol is said to be efficient. As the complexity of the protocol increases, the effort required for collaborators to set common grounding and SA increases.

In a remote-helper and worker protocol, the helper provides step-by-step instructions to the worker to complete the task. If the worker responds to the helper's instructions with an acknowledgment of understanding or positive action, then the helper knows that the worker has understood his instruction, and when the helper moves to new instruction, then the worker infers that the helper knows that he understood the helper's instruction. Here the helper extends his knowledge to the worker and establishes a mutual belief with the worker; then common ground is said to be drawn between the helper and worker. If the worker responds with a question or performs a wrong action following an instruction from the helper, then the helper knows that the worker did not understand his instruction. Subsequently, when the helper reiterates the instruction, then the worker infers that the helper knows that he did not understand the helper's instruction. Here the helper's instruction fails to establish the grounding with the worker in the initial attempt, so the helper shares more information with the worker to establish the grounding. With this, the effort by helper and worker to establish the grounding increases, thereby increasing the complexity of the grounding protocol. The complexity of a protocol also depends on the medium of communication used for the collaboration. If the mediating communication system does not provide SA and establish the grounding, the number of requests and responses between the collaborators increases, thereby increasing the complexity of the protocol.

In the following section, we discuss the existing methodologies for measuring the performance of a helper-worker collaboration. We propose a new methodology for measuring complexity and minimal complexity of a protocol. One can use the minimal protocol complexity as a baseline for identifying the extent to which a given helper-worker protocol achieves minimal complexity.

## 5.1 Existing Methodology for Analyzing Efficiency of the Remote Collaboration

**Table 2: Existing methodologies for analyzing the efficiency of the remote collaboration**

| Existing Methodology | Drawbacks of Existing Methodology |
|---|---|
| Ranjan et al. [16]<br><br>On a video-mediated collaborative task between geographically distributed groups, Ranjan et al. utilized task completion time, the number of critical errors made by participants, and self-reported effectiveness as metrics for analyzing the performance of the collaborative task. | Existing methodologies focus on studying the performance of the mediating systems and analyzing the impact of different sources of information on the collaboration rather than analyzing the performance (i.e., success or failure) of the collaboration itself. Therefore, these methods cannot be used for analyzing the complexity of a helper-worker protocol. We propose a methodology to analyze the performance of the collaboration itself. The methodology is explained in the rest of section 5. Our proposed method can be used to understand the complexity of the protocol itself. |
| Kraut et al. [13]<br><br>Kraut et al. use the number of tasks completed, completion time, repair quality, rating of work quality, participant's communication on a 5-point scale, conversational coding, and the number of deictic references in the conversation as metrics. | |
| Other works by [15][17][18][21][23][25]<br><br>Many other researchers also used task completion times, quality of task, mean number of words per task, number of utterances, number of mistakes, and user rating as metrics to measure the efficiency of the remote collaboration system. | |

## 5.2 Modeling a Task Using Graph Structures

Robert E. Wood [26] presents a theoretical model to calculate the complexity of a task using three components of a task: products, acts, and information-cues. He describes products as entities created or produced by behaviors; acts as the pattern of behaviors with some identifiable purpose or direction; and information-cues as pieces of information about the attributes of stimulus objects upon which an individual can base the judgments he or she is required to make during the performance of a task. Haerem et al. [27] have developed an extended concept of task complexity. They retain Woods' concept of representing a task as actions and information-cues and conceptualize a new task complexity of multiple actors working together. Pentland models tasks as a network of events, where actions by actors are events, and these events generate information-cues. Representing a task as events and information-cues and modeling it as a network helps in identifying the complexity in various methods.

Following Wood's [26] representation of a task, we propose a methodology to model a helper-worker protocol. We represent a helper-worker protocol of a physical task as a network of events where an event is an utterance or an action by a helper or worker and events generate information-cues that may help another event to occur.

## 5.3 Quantitative Measures of a Graph Complexity

Bonchev [28] states that one can obtain an appropriate complexity measure of a graph by using the vertex degree magnitude-based information content. Bonchev [29] developed a method to quantify the topological complexity of a directed biological network; he utilized Shannon's information theory [34] to estimate complexity based on entropy. He assumes the distribution of N elements in k groups as {N1, N2, N3, … Nk} and the probability for a randomly chosen element of the set to belong to the group 'i' is Pi=Ni/N. Shannon's entropy H of probability distribution {P1, P2, P3, … Pk} is defined as,

$$H́ = -\sum_{i=1}^{K} P_i \log_2 P_i = -\sum_{i=1}^{K} \frac{N_i}{N} \log_2 \frac{N_i}{N} \text{ bits/element} \quad (1)$$

Here log is taken as base 2 to calculate entropy in bits. The entropy of all the elements is given by,

$$H = N \times H́ = N \times \left(-\sum_{i=1}^{K} \frac{N_i}{N} \log_2 \frac{N_i}{N}\right) \quad (2)$$

$$H = N \log_2 N - \sum_{i=1}^{K} N_i \log_2 N_i \quad (3)$$

As per Bonchev, the entropy of a structure H is maximized when the second term in the above equation is zero; this corresponds to distributing system elements into groups of one element each. The information content of a graph is the difference between the maximum entropy $H_{max}$ and the value H of the system entropy.

$$I = H_{max} - H = \sum_{i=1}^{K} N_i \log_2 N_i \quad (4)$$

Bonchev [28] states that the information content of a structure can be used to calculate its topological complexity. This equation, derived for information content, can be used for calculating the complexity of a structure.

For utilizing Bonchev's structural complexity theory in calculating the complexity of a protocol, representation of a protocol as a structure is crucial. In the next section, we discuss the methods of representing a protocol as a graph structure.

## 5.4 Modeling a Helper-Worker Protocol

To quantify the complexity of a protocol, we must first understand its information content. Representing a protocol as a graph structure helps to measure its information-content. We represent a protocol as a graph by presenting events and information-cues as vertices and edges respectively.

In a collaboration, the worker responds to a helper's instruction with either action, question, or acknowledgment. We represent these requests and responses as events and the information signals provided by them as edges. In a collaboration, every request acts as a source of information for the response and every response acts as a source of information for the next request.

Common grounding is crucial to collaboration. A helper follows a step-by-step process for accomplishing a task. Each step in this process starts with an instruction from the helper and ends with the successful execution of that instruction by the worker. The complexity of the interaction depends on how well the helper and worker can establish the grounding. An instruction by a helper leads to further events; the complete set of these events and the information-cues that are required either to ground the helper and worker or to complete the helper's instruction are defined as an atomic task. An atomic task involves successful grounding and execution of a helper instruction. The rest of this section discusses the modeling of different types of atomic tasks.

When a helper provides an instruction to the worker to act and the worker executes the instruction correctly, the helper knows that the worker understood his instruction. Here the grounding between the collaborators is achieved with one information cue. This atomic task is minimally complex. Figure 5 shows a graph representation of this type of atomic task.

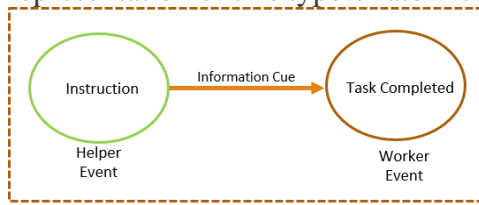

**Figure 5: Type I atomic task**

When a helper provides an instruction to the worker and the worker performs a wrong action, it indicates that the worker misunderstood the instruction and the helper did not establish a common ground with the worker. Then the helper provides additional information and tries to set the grounding with the worker. This atomic task is more complex compared to the Type I atomic task. Figure 6 shows a graph representation of this atomic task.

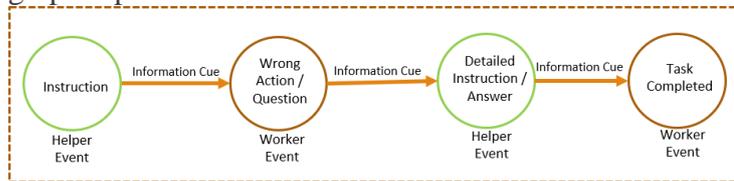

**Figure 6: Type II atomic task**

When a helper provides an instruction and the worker completes it partially, it indicates that the helper established grounding with the worker, but either the worker or the helper is not sure of the extent to which the action should be performed. If the helper is not sure of the extent of action required, after he gains visual information of the worker's action, he reiterates the same instruction to convey the message to continue the current action. If the worker is not sure to what extent he should act, the helper, after fetching the status of the current action, reiterates the same instruction. Figure 7 shows a graph representation of this atomic task.

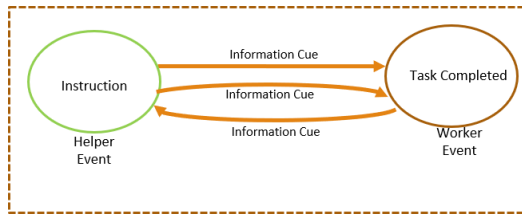

**Figure 7: Type III atomic task**

When a helper provides lengthy information about the objects or the process, he provides it in installments and tries to establish common ground with the worker for every installment. When the helper receives an acknowledgment for the first installment, then the helper knows that the worker understood his first portion of information and proceeds with the second installment of information. For each block of information, the worker may immediately ground with the helper by providing an acknowledgment or requesting additional information. Figure 8 provides the graph representation for this type of atomic task.

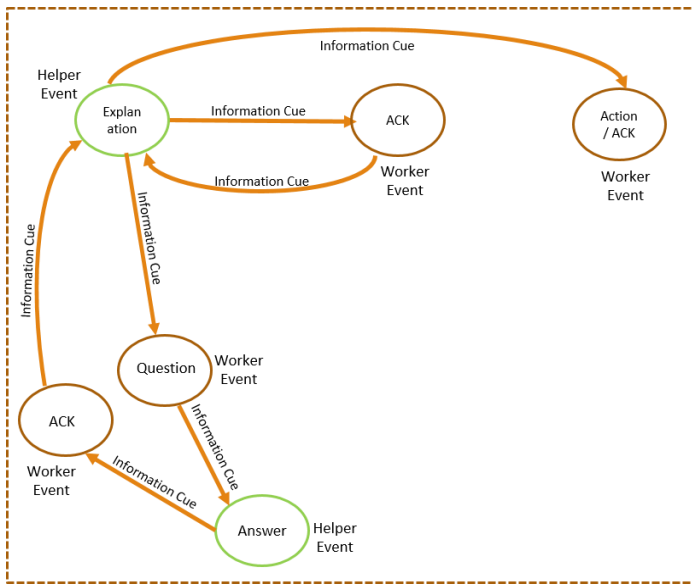

**Figure 8: Type IV atomic task**

When a prolonged action is to be performed, the worker grounds with the helper and starts executing the task. During this continuous action, the worker may unknowingly slide out of the right path; the helper provides mid-action information to correct the worker's actions. Alternatively, the helper provides the mid-action information to pass the message to the worker that he is on the right track or to convey the status of the task. Figure 9 shows the graph representation of this type of atomic task.

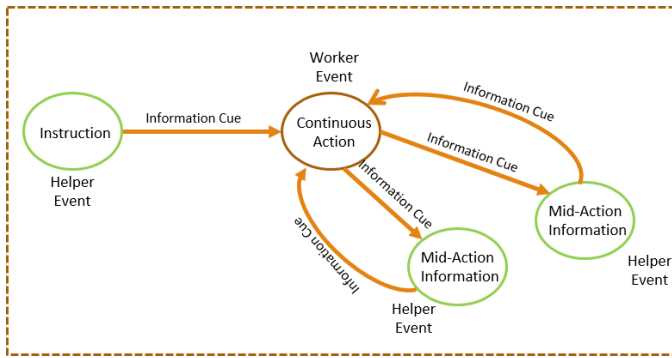

**Figure 9: Type V atomic task**

## 5.5 Modeling Complexity of a Protocol

A protocol is a combination of atomic tasks; the information content of a graph provides its structural complexity. When one represents a protocol as a graph structure and measures its structural complexity, it provides the complexity of the protocol. The term degree of a vertex provides the number of edges associated with it. For a protocol structure, the degree represents the information signals generated or processed by an event. To calculate the protocol complexity, we calculate the complexity of each event by considering the information-cues associated with it. Summing the complexities of all vertices provides the complexity of the overall graph, thereby providing the complexity of the protocol. The information-cues of an event are the sum of the information generated by it and the number of information-cues processed by it.

$IC_i = InDegree(node_i) + OutDegree(node_i)$  (5)

The complexity of an event is given by

$EC = IC \log_2 IC$  (6)

Complexity of a protocol is given by

$PCI = EC_1 + EC_2 + EC_3 + \cdots\cdots\cdots + EC_k$  (7)

$PCI = IC_1 \log_2 IC_1 + IC_2 \log_2 IC_2 + IC_3 \log_2 IC_3 + \cdots\cdots + IC_k \log_2 IC_k$  (8)

$PCI = \sum_{i=1}^{K} IC_i \log_2 IC_i$  (9)

where PCI is the protocol complexity index; K is the total number of vertices or events in the protocol; ICi is the number of information-cues associated with an event or vertex i.

Figure 10 shows the sensitivity of the protocol complexity index PCI to changes in the information-cues IC and events K. Both the dependent variable events and information-cues start with a value 2 (there should be at least two events to an information cue). The X-axis shows the percentage increase of the cues and events and the Y-axis shows the increase in protocol complexity index to changes in information-cues and events. For analyzing the sensitivity of the complexity index to information-cues, the number of events is considered constant; similarly, information-cues are considered constant for analyzing the sensitivity of the complexity index to events

Moreover, for the purpose of sensitivity analysis of PCI, all the events are assumed to have the same degree of information-cues. With this, the protocol complexity equation reduces to the following equation.

$PCI = K * IC_i \log_2 IC_i$  (10)

The graph in Figure 10 shows that the complexity PCI is more sensitive to changes in information-cues than to changes in events. The complexity of approximately 4800 bits is achieved for 1000% increase in information-cues, where it took a 4800% increase in events to reach the same complexity. This sensitivity analysis conveys that if a protocol is designed to minimize the number of information-cues for each event, it can reduce the complexity.

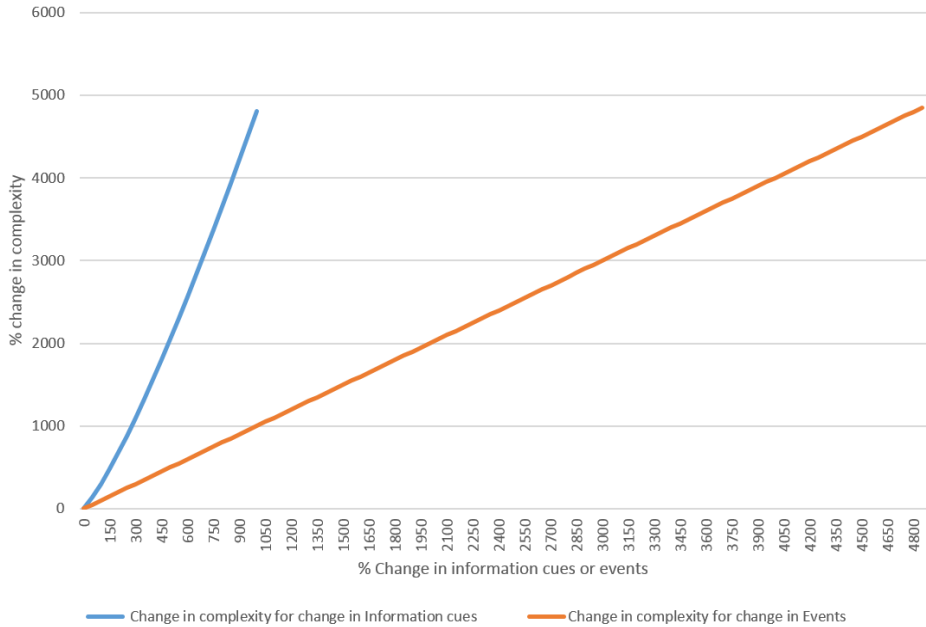

**Figure 10: Sensitive analysis of protocol complexity index to changes in the number of information-cues (IC) and the number of events (K)**

We observed similar percentage changes in the complexity index for the percentage change in information-cues for different values of the constant events. Similarly, we observed similar percentage changes in the complexity index for the percentage change in the events for different values of the constant information-cues.

### 5.5.1 Performance Index of a Collaboration

PCI provides a metric to analyze the efficiency of a protocol and therefore the collaboration. The rate at which the protocol is executed determines the overall performance of the collaboration. The performance of a collaboration is the ratio between protocol complexity and time taken by the collaboration.

$$\text{PerformanceIndex } P_I = \frac{\text{ProtocolComplexityIndex}}{\text{Time}} \quad (11)$$

$$P_I = \frac{\sum_{i=1}^{K} IC_i \log_2 IC_i}{\text{Time}} \quad (12)$$

where PI is the performance index of the collaboration, time is the time taken by the collaboration, and $IC_i$ is the number of information-cues associated with an event i.

## 5.6 Mathematical Minimization of a Protocol

When a worker requests help on a task, he wants to complete the job with the least collaboration effort possible. Achieving the least collaboration effort means performing the job with a minimally complex protocol. Obtaining minimal complexity for a protocol means the helper should establish common ground with the worker and the worker should execute the helper's instructions with the fewest possible events and information-cues. To judge if the helper and worker protocol is minimally complex, one needs a baseline for comparison. This baseline is a theoretical value satisfying the constraints of a given protocol.

The complexity of a protocol depends on multiple factors: the efficiency of the mediating communication system, prior knowledge of the worker on the task, and knowledge and capability of the helper over remote collaboration. So, the protocol complexity is different for each worker and cannot be compared across workers. We developed a methodology for calculating the minimal complexity of a protocol. Using this method, one can deduce a minimal complexity value for each worker. Moreover, this can act as a baseline for judging if the helper and worker achieved a minimally complex protocol.

Our methodology was developed by performing constrained mathematical optimization of the protocol complexity equation. We modeled the minimal complexity equation from the atomic task level and considered below constraints in an atomic task to model a minimal protocol complexity equation.

- Each atomic task should have only one helper event
- An atomic task is minimally complex when there is only one helper event in it. An atomic task is a successful grounding and execution of an instruction from the helper. When an instruction from a helper is successfully grounded or executed by the worker, there is no need for additional information or instructions from the helper. Then the task is minimally complex.
- Two lower bounds are considered to avoid off shooting of the optimal complexity values
- A non-root helper event should have at least one in-degree
- A helper event should have information-cues that are at least twice the number of non-leaf worker nodes (n) and at least one information cue associated with the last leaf node in the atomic task.
- Each worker event should have one input and one output information cue.
- Indegree (Helper_Event)>=1
- Degree (Helper_Event) with non-leaf worker nodes = 2 * n; where n = N-1
- Degree (Helper_Event) with leaf worker node = 1
- Total Degree (Helper_Event) >= Indegree (Helper_Event) + Degree (Helper_Event) with non-leaf worker nodes + Degree (Helper_Event) with leaf worker nodes

$TotalDegree(Helper\_Event) \geq 1+2*(N-1)+1$ \hfill (13)

$Degree(Helper\_Event) \geq 2*N$ \hfill (14)

$Degree(Worker\_Event) \geq 2$ \hfill (15)

The helper node in the first atomic task does not have an in-degree, so we are considering the first atomic task and the rest of M atomic tasks separately. Performing constrained minimization of the protocol complexity equation (PCI) obtains the following optimal complexity equation:

$$\sum_{i=1}^{M}[2N_i \log_2 2N_i]+2K+(2L-1)\log_2(2L-1) \quad (16)$$

where Ni is the number of worker nodes connected to first helper node in atomic task i, K is the total number of worker nodes in a protocol, L is the number of worker nodes associated with the root helper node and M is the number of atomic tasks excluding the atomic task with root helper node.

# 6 Experiments and Results

## 6.1 Vehicle Protocol

The statistics of rescued drivers with vehicle breakdowns provide a clear indication of the portion of licensed drivers with an awareness of auto problems and repairs [30]. Sometimes the availability of experts does not meet the driver's needs, for example when an expert is needed during odd hours, inclement weather or in remote locations. Not all vehicle breakdowns require a towing to a repair shop; many drivers can fix minor repairs with the required expertise.

AAA reports tire-related issues being among the major causes of vehicle breakdowns [31]. Replacing a flat tire is not among the skills every driver has, nor should changing a tire be a trial and error process, as incorrect actions may cause damages both to the driver and vehicle. One of the crucial components in replacing a vehicle tire is positioning a jack securely and elevating the car. Cars have different locations for positioning the jack, called pinch points; this may make it difficult for the driver to identify where to position the jack. With the help of CARE, a driver can get technical guidance from a remotely-located mechanic and fix the problem.

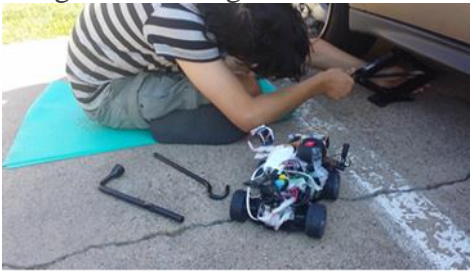

**Figure 11: A driver positioning a jack for changing the tire with the help of a remote-helper**

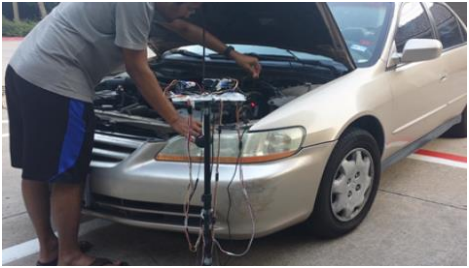

**Figure 12: A driver inspecting the engine-oil level with the help of a remote-helper**

### 6.1.1 Car Elevation Protocol

We perform car elevation task experiments with multiple subjects to study the remote-helper and worker protocol complexities. We compare the achieved protocol complexities with a baseline complexity index to analyze whether a remote-helper can guide a worker on a car elevation task with optimal protocol complexity. In this section, we explain the car elevation protocol and provide a snapshot of the car elevation process.

Replacing a vehicle's flat tire involves multiple steps and requires knowledge of the vehicle, tools, and process.

- Knowledge about the vehicle: The jack (used to elevate the car) should be placed under the vehicle at an appropriate place called a pinch point. Pinch points vary by vehicle. A worker should have knowledge of the characteristics of the vehicle.
- Tools: There can be a variety of instruments each car carries; vendors sell different sets of tools along with a vehicle. A worker should have knowledge of what tools are available and how to use them.
- Process: There are multiple steps involved in elevating the car; a worker should know the efficient way to proceed and the appropriate amount of elevation required.

Using CARE, the remote-helper fetches the required information regarding the worker's environment, and with the help of voice communication and gestures, the helper provides step-by-step instructions to the worker. The sets of requests and responses between the helper and worker are called a protocol. This protocol is crucial to the success of a collaboration. An optimally complex protocol means a highly efficient protocol.

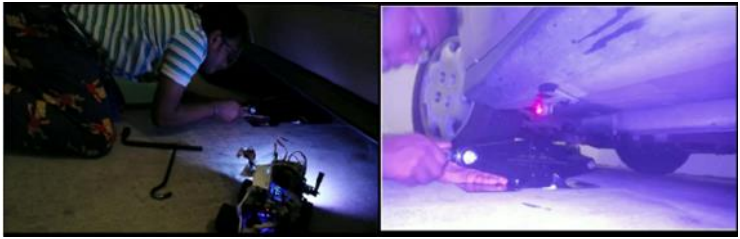

**Figure 13: Worker positioning the jack under a pinch point and helper monitoring it; the worker's environment is on the left and the helper's view is on the right**

*Modeling the Car Elevation Protocol and Analyzing the Efficiency Using Protocol Complexity Theory*

We can represent the car elevation protocol as discussed in section 5, using a graph structure. An example graph representation of car elevation protocol is provided in appendix A. The complexity of the car elevation protocol depends on multiple factors: the efficiency of the mediating device, the worker's prior knowledge on the task, and the worker's comprehensive capability. Using equation 16, one can deduce an optimal complexity value for each worker.

*Results*

To observe the remote-helper and worker protocol complexity and to analyze whether the remote-helper can guide a worker with optimal protocol complexity, experiments were conducted using multiple subjects. In this section we present the demographic data of the subjects who participated in the experiments, calculations of the complexity index of different car elevation protocols, and the comparison of the archived protocol complexity values with optimal complexity index, to analyze the satisfaction of the protocols to the objective.

The subjects who participated in the experiments had diverse (0 to 8 years) driving experience. The subject pool included drivers with and without knowledge of car elevation using a jack. Table 3, Table 4, and Table 5 show the demographic data of the subjects by gender, year of experience, and prior knowledge of car elevation respectively.

**Table 3: Demographics of subjects for car elevation protocol**

| Gender | Age Group | Average and Standard Deviation | Number of subjects |
|--------|-----------|-------------------------------|--------------------|
| Female | 21 - 27 | 23.16 ± 1.48 | 26 |
| Male | 21 - 31 | 23.48 ± 2.18 | 33 |

**Table 4: Prior knowledge of subjects on car elevation for tire changing**

| Gender | 0 years | <1 year | 1-3 years | 4-6 years | >6 years |
|---|---|---|---|---|---|
| Female | 10 | 11 | 4 | 0 | 1 |
| Male | 4 | 11 | 9 | 5 | 4 |

**Table 5: Prior knowledge of subjects on car elevation for tire changing**

| Gender | Have prior knowledge of car elevation | Do not have prior knowledge of car elevation |
|---|---|---|
| Female | 3 | 23 |
| Male | 12 | 21 |

*Relationship Between Car Elevation Protocol Complexity and Time*

The car elevation protocol experiment was conducted with 59 subjects. We received complexity values ranging from 77.916 to 284.012 bits and the time taken to accomplish the task ranged from 122 to 713 seconds. We observed that the time taken to accomplish a task increased linearly with the complexity of the task, as shown in Figure 14. We then performed regression analysis to find the strength of the relationship between PCI in bits and time in seconds and we obtained the P-value of $1.389 * 10^{-5}$ and a standard error of 94.65 seconds. The P-value indicates a strong relationship between time (T) and the protocol complexity index (PCI). This proves that the Protocol Complexity Equation can be used for calculating the protocol complexity of a vehicle protocol.

In the scatter plot shown in Figure 14, the majority of the data points are in the cluster of complexity range 112 - 198 bits and time range 244 - 485 seconds. A few data values are spread away from this cluster with the protocol complexity values higher than normal for a few data points and time values higher for a few other points.

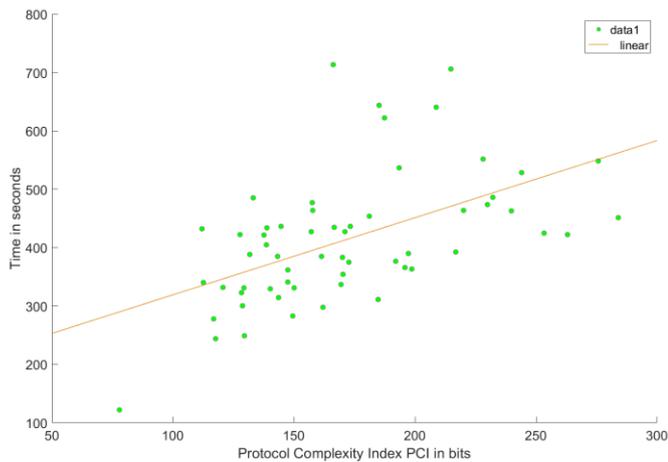

**Figure 14: Relationship between PCI and Time consumed by the protocol**

We used the optimal complexity index as a baseline for comparing protocol complexity values. The index indicates whether the achieved protocol complexity value from the experiment is optimal. Then, in order to study the impact of driving experience on protocol complexity, we included the driving experience. In Figure 15, blue bars indicate experimental average complexity and orange bars indicate average optimal complexity values. Figure 15 shows that the difference between experimental and optimal complexity values falls with the increase in the driving experience. Furthermore, the graph does not show a relation between driving experience and experimental protocol complexity values. Therefore, we can conclude that car elevation protocol is independent of the subject's driving experience.

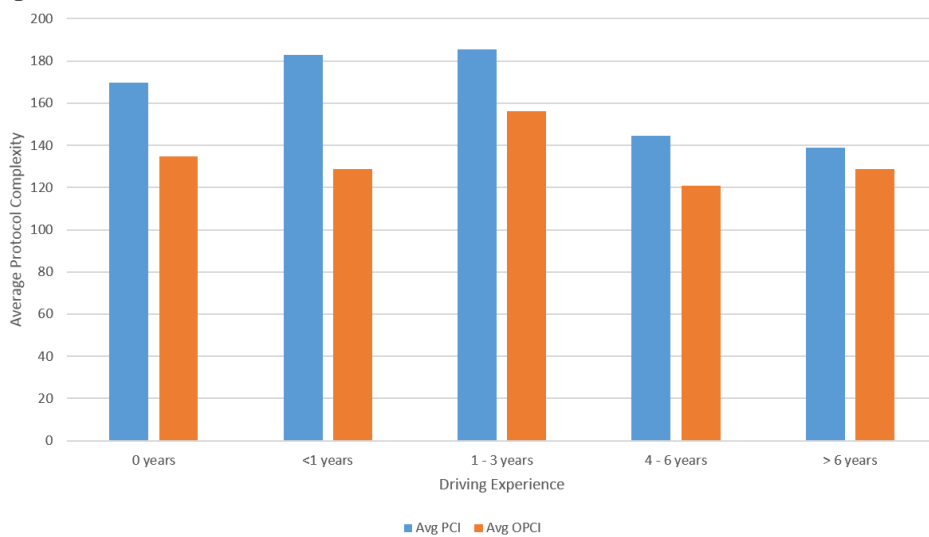

**Figure 15: Relationship between a subject's driving experience, average protocol complexity, and average optimal protocol complexity**

To justify that the data set used for analyzing car elevation protocol complexity is sufficient for the claims we make on the relation between achieved and optimal protocol complexities, we consider a cumulative average of protocol complexity index values. Figure 16 shows that after 40 subjects the average complexity index remains relatively constant.

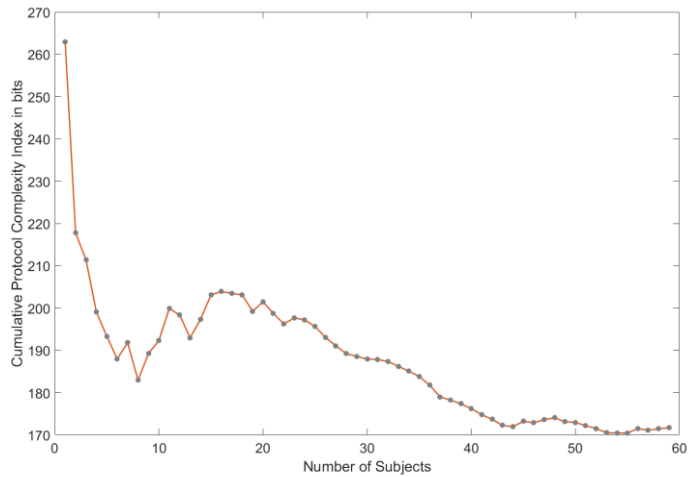

**Figure 16: Cumulative average of PCI among the range of subjects**

We conducted a survey after each experiment to get feedback on the collaboration. We showed each of the subjects the same YouTube video on car jack elevation. Our survey then requested the subjects to rate the helper-worker protocol and the YouTube video in eight areas, focusing on the quality of the information and the subject's satisfaction on task completion. The survey results show that the subjects believed that they used less mental effort and were less frustrated and confused while performing the task with CARE than they would have been having watched only the YouTube video. Figure 17 shows the questions we asked the subjects and how subjects rated the methodologies.

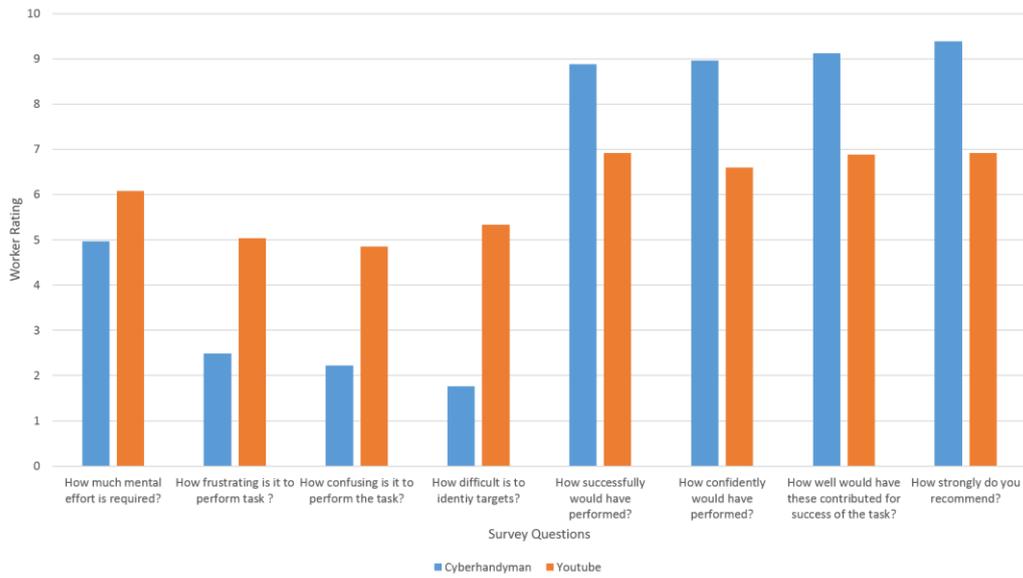

**Figure 17: Post vehicle protocol survey results**

### 6.1.2 Underneath Vehicle Inspection

Security inspection of vehicles has become a security routine at many hotels, airports, shopping malls, and other venues in which large populations gather. In a security inspection, the inspector examines vehicles thoroughly for any hidden explosives or weapons. With CARE, a remote device can inspect the vehicles for threats, keeping the inspector at a safe distance from the vehicles. The device can move around under the parked vehicles and make a thorough inspection largely without delaying traffic. If equipped with image detection capabilities the device can be used to identify weapons and generate an alarm.

In addition to security inspection, a remote-helper can be useful for inspecting under vehicles that may have sustained any type of damage. The below pictures show a helper's view when using CARE for underneath vehicle inspection.

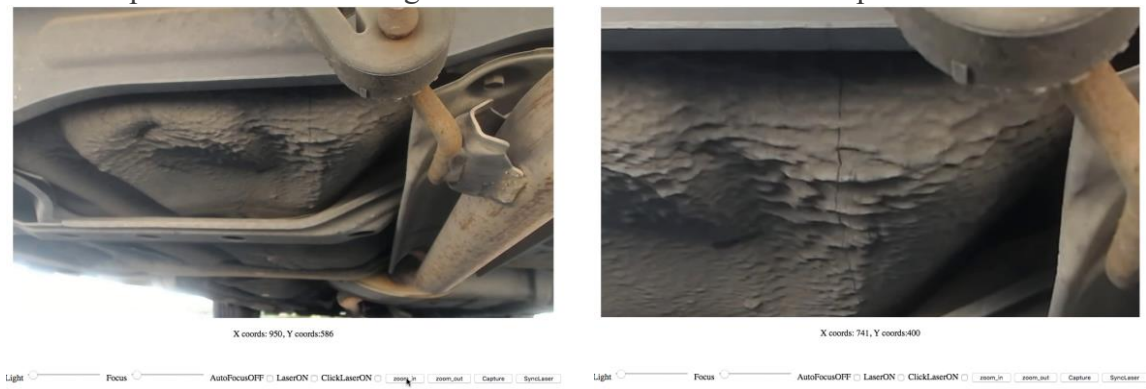

(a) Normal view of the cracks  (b) Close up view of the cracks

**Figure 18: Under-vehicle inspection of cracks**

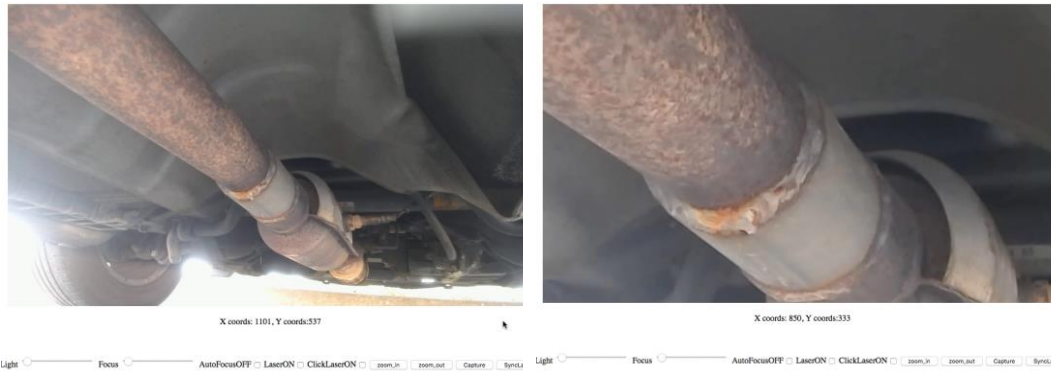

(a) A normal view of the rusted part  (b) Close up view of the rusted part

Figure 19: Under-vehicle inspection for rust

### 6.1.3 Tire Depth Analysis

Appropriate tire depth is crucial for a vehicle's safety. Worn out tires are more prone to blasts causing crashes on roads. Tire treads may wear unevenly when the wheels are out of alignment; the inner side of the tires wears out more than the outer side, and often this goes unnoticed. Using CARE, a remote mechanic can analyze the depth of the tire tread. With edge detected images of the tires, the remote expert can effectively analyze the tread quality.

We performed Canny's edge detection over both worn-out and new tires. The tires can be differentiated based on their tread quality in the edge detected images shown in Figure 20. The tire with good tread has edges that are visible, and the tire with worn-out tread has barely visible edges.

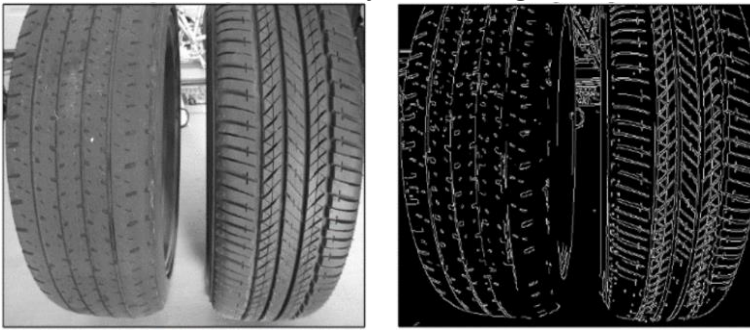

(a) Original image      (b) Edge image

Figure 20: Edge detection of tires [35]

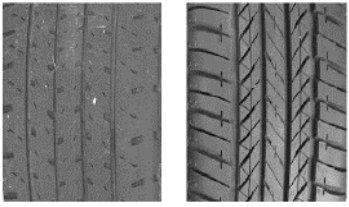

**Figure 21: Portion of tires used for pixel intensity calculation**

We also analyzed pixel intensities of a portion of the two tires shown in Figure 21. As observed in the pixel intensity graphs in Figure 22, the new tread has more pixels within the 0-100 range compared to worn-out tread.

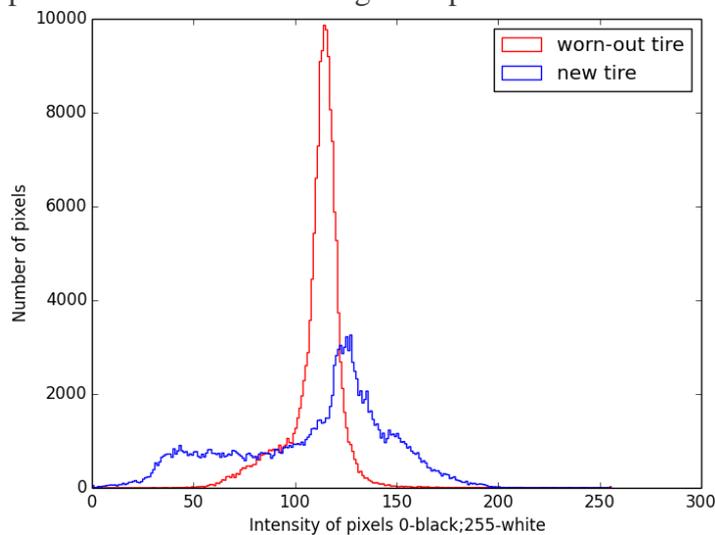

**Figure 22: Pixel intensities vs. number of pixels**

## 6.2 Home Repair Protocol

Per the 'Handyman Services and Pricing Guide' article by homeadvisor.com [32], homeowners pay between $165 and $611 for hiring a handyman; the national average is $388 per handyman visit. On average, a homeowner in the United States spent $1,105 on home maintenance in the year 2018 and $7,560 on home improvement [33]. When homeowners perform repairs by themselves, they may cause further damages and spend even more on repairs because they lack the knowledge to properly complete the tasks. Furthermore, when repair projects are put off due to expensive repairs, the problem may expand and cause further damages. Attending to a home repair immediately can save the homeowner money, even if the immediate repair action is simply "first-aid" while awaiting the expert. Often only an experienced professional has the skill set necessary for a repair.

A home repair may involve multiple steps, which include troubleshooting and identification of cause for the problem, identification of required materials and tools, and an efficient way to fix the problem. False assessment or an improper approach may lead to an increase in both the cost and complexity of the repair.

There are different categories of home repair that require a range of skills from basic to expert level. The categories include plumbing, electricity, carpentry, and appliance repairs. Some repairs cannot wait; for example, plumbing leaks that seem minor can quickly transform into a major problem damaging the walls and floors. Fixing plumbing leaks can save a homeowner from incurring unnecessarily high water bills. With an expert's guidance, some homeowners may be able to perform necessary repair tasks.

Be it technical assistance on a temporary fix, performing a repair, troubleshooting, or identifying required materials, a homeowner can obtain the service from an expert anytime from anywhere with the help of CARE.

### 6.2.1 Faucet Repair Protocol

We performed faucet repair experiments with multiple subjects to study the remote-helper and worker protocol complexities. We compare the achieved protocol complexities with a base-line complexity index to analyze whether a remote-helper can guide a worker on a faucet repair task with optimal protocol complexity. In this section, we explain the faucet repair protocol, provide a snapshot of the faucet repair process, and discusses the methodology to analyze the complexity of the collaboration protocol.

Fixing a leaky faucet involves multiple steps; if one performs this task without prior knowledge, he may cause damage to the faucet causing more water to leak. The worker requires knowledge of tools, troubleshooting the problem, identifying the parts to be replaced, removing the parts, and placing the parts back in the correct order and orientation.

Using CARE, the remote-helper can acquire the necessary information about the worker's environment and guide the worker through the repair using voice communication and gestures.

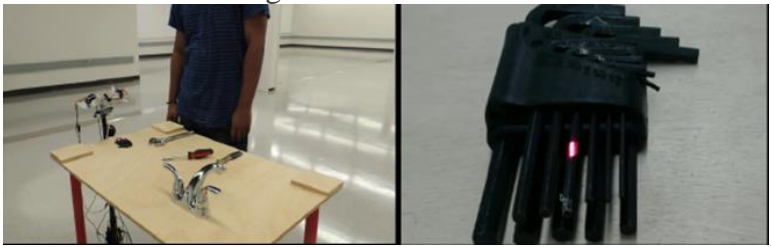

**Figure 23: Remote-helper inspecting different sizes of hex keys and instructing the worker to use a specific key**

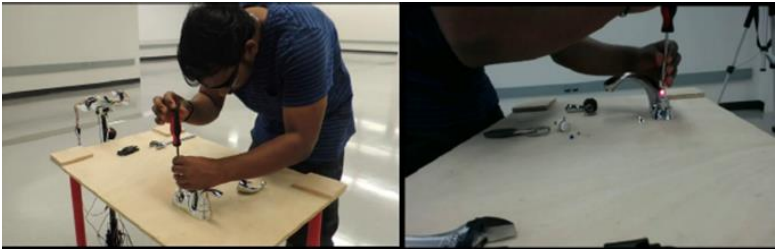

**Figure 24: Worker unscrewing a screw using a screwdriver with instruction from the remote-helper**

*Modeling and Analyzing Complexity of Faucet Repair Protocol*

Using the proposed protocol complexity theory, one can analyze the complexity of the faucet repair protocol. The complexity of the instruction and the response is minimal when the worker understands and executes the instruction correctly without having further questions.

The complexity of the faucet repair protocol depends on multiple factors, such as the efficiency of the tools, prior knowledge of the worker, and the comprehensive capability of the worker. To find the optimal possible complexity for a protocol, the complexity minimization equation can be used.

*Results*

To study the remote-helper and worker protocol complexities and to analyze if the remote-helper can guide a worker with optimal protocol complexity, we performed several experiments with multiple subjects. The demographics and prior knowledge of subjects for the faucet repair protocol are shown in Table 6 and Table 7.

**Table 6: Demographics of subjects for faucet repair protocol**

| Gender | Age Group | Average and Standard Deviation | Number of subjects |
| --- | --- | --- | --- |
| Female | 21 - 24 | 22.71 ± 1.16 | 7 |
| Male | 21 - 29 | 23.39 ± 2.34 | 23 |

**Table 7: Prior knowledge of subjects on faucet repair**

| Gender | Have prior knowledge of faucet repair | Do not have prior knowledge of faucet repair |
| --- | --- | --- |
| Female | 0 | 7 |
| Male | 6 | 17 |

We performed the faucet repair protocol experiment with 30 subjects and obtained a complexity value ranging between 171.97 and 466.35 bits and time ranging between 404 and 865 seconds. To find the applicability of the protocol complexity to the faucet repair protocol, we plot the relationship between protocol complexity and the time taken for the protocol. As shown in Figure 25, we discovered that the time taken for the protocol increases linearly with its complexity. To find the strength of the relation between PCI in bits and time in seconds, we perform regression analysis and obtain the regression equation below with a P-value of $2.379 * 10^{-4}$ and a standard error of 105.66 seconds. The P-value indicates a strong relation between Time and PCI. This proves that the protocol complexity equation can be used for calculating the protocol complexity of a faucet repair protocol.

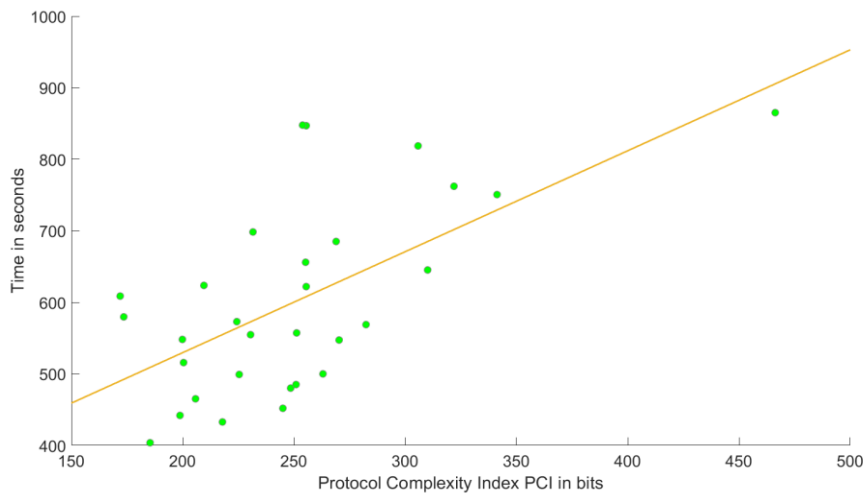

**Figure 25: Relationship between PCI and Time consumed by the protocol**

The optimal complexity index (OPCI) acts as a baseline for comparing protocol complexity values and validating if a given protocol complexity is optimal. Figure 26 shows a comparison between the achieved and calculated protocol complexity values. In the graph in Figure 26, the y-axis is the complexity values in bits; blue bars indicate the experimental complexity and orange bars indicate optimal complexity values.

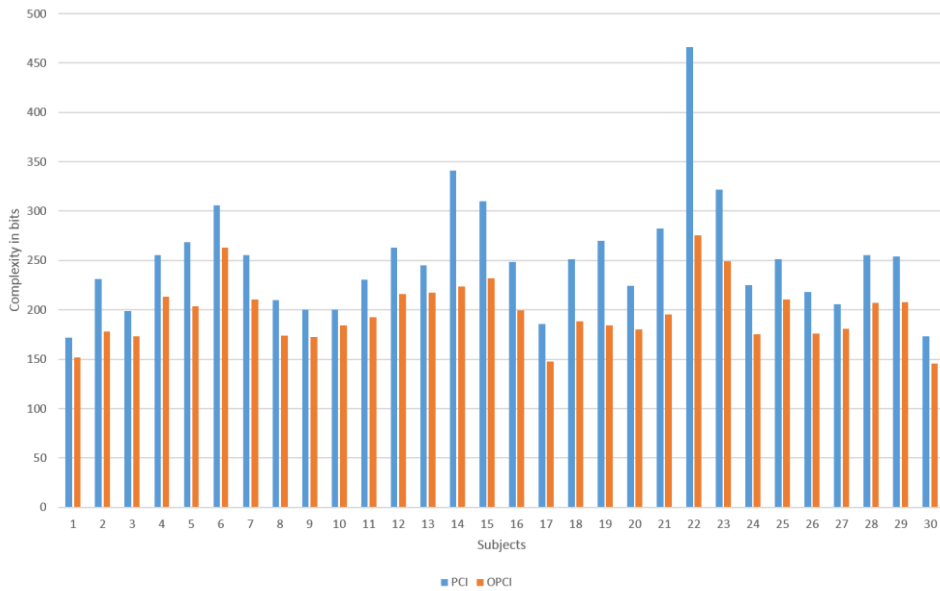

**Figure 26: Relationship between PCI and OPCI for each protocol**

We conducted an eight-question survey after the faucet repair experiment to get feedback from the subjects on the helper-worker collaboration. We showed a YouTube video on faucet repair to all the subjects and requested the subjects to rate both the helper-worker protocol and the YouTube video. The questions on the survey focused on the quality of the information and a subject's satisfaction on task completion. The survey results shown in Figure 27 depict that subjects believed CARE can enable them to accomplish the task more effectively than watching a YouTube video for faucet repair protocol would.

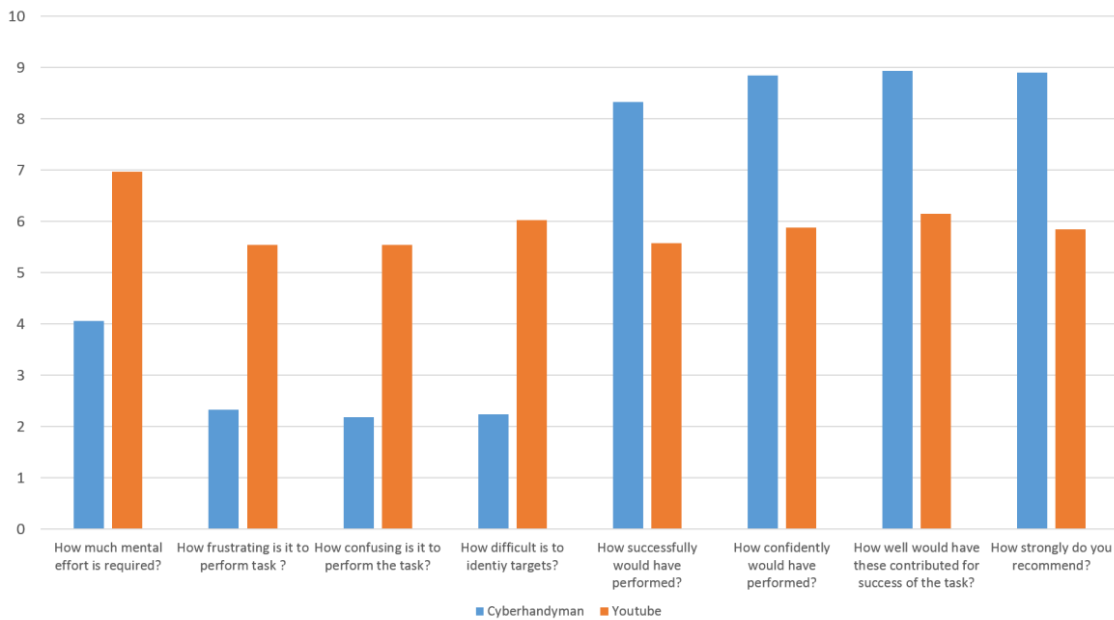

**Figure 27: Post-task survey results for faucet repair protocol**

# 7  Conclusion

This paper introduces a solution to the problem of the collaboration of geographically separated people on physical tasks. We discussed existing theories and impacts of different sources of information on the performance of collaboration. Based on existing theories of common grounding and SA, we developed a communication system to meet the requirements of remote collaboration.

Using actions and utterances as events and the messages conveyed by these events as information-cues, we represented the protocol as a structure and developed a methodology to measure the complexity and minimal complexity of a protocol. Data collection of helper and worker protocols of car elevation and faucet repair protocols proved that a remote helper would be able to guide a worker on a physical task with minimal difficulty.

We represented the events and information-cues as vertices and edges in graphs and modeled a protocol as a graphical structure. Complexity of the protocol is obtained by calculating the complexity of the graphical structure. The complexity of a structure is the information content of the structure. Furthermore, we presented a model for calculating the optimal complexity of a protocol structure to analyze whether the helper and worker achieved the optimal complexity of the protocol.

We conducted remote collaboration experiments with fifty-nine subjects over a car elevation task and with thirty subjects on a home repair task. To understand the applicability of the provided methodology in measuring the complexity of a car elevation and home repair protocol, we provided a relation between time and measured protocol complexity for each task. The regression analysis of both tasks showed a strong relationship between the two variables indicating the applicability of the equation for those protocols. The majority of subjects in both tasks obtained a complexity value closer to the minimal complexity. The surveys from all 89 participants on the effectiveness of the remote-helper versus a YouTube video on the task showed that the users are more comfortable with a remote-helper guiding them through the task rather than watching a video performing the task. All these results strengthen the hypothesis that with the help of a remote-helper and an efficient collaboration system, workers can execute a physical task with minimal difficulty.

# A APPENDEX

The graph representation of car elevation protocol is shown in Figure 28 below. The structural representation of car elevation protocol divides the protocol into atomic tasks. Atomic tasks are represented by dotted lines.

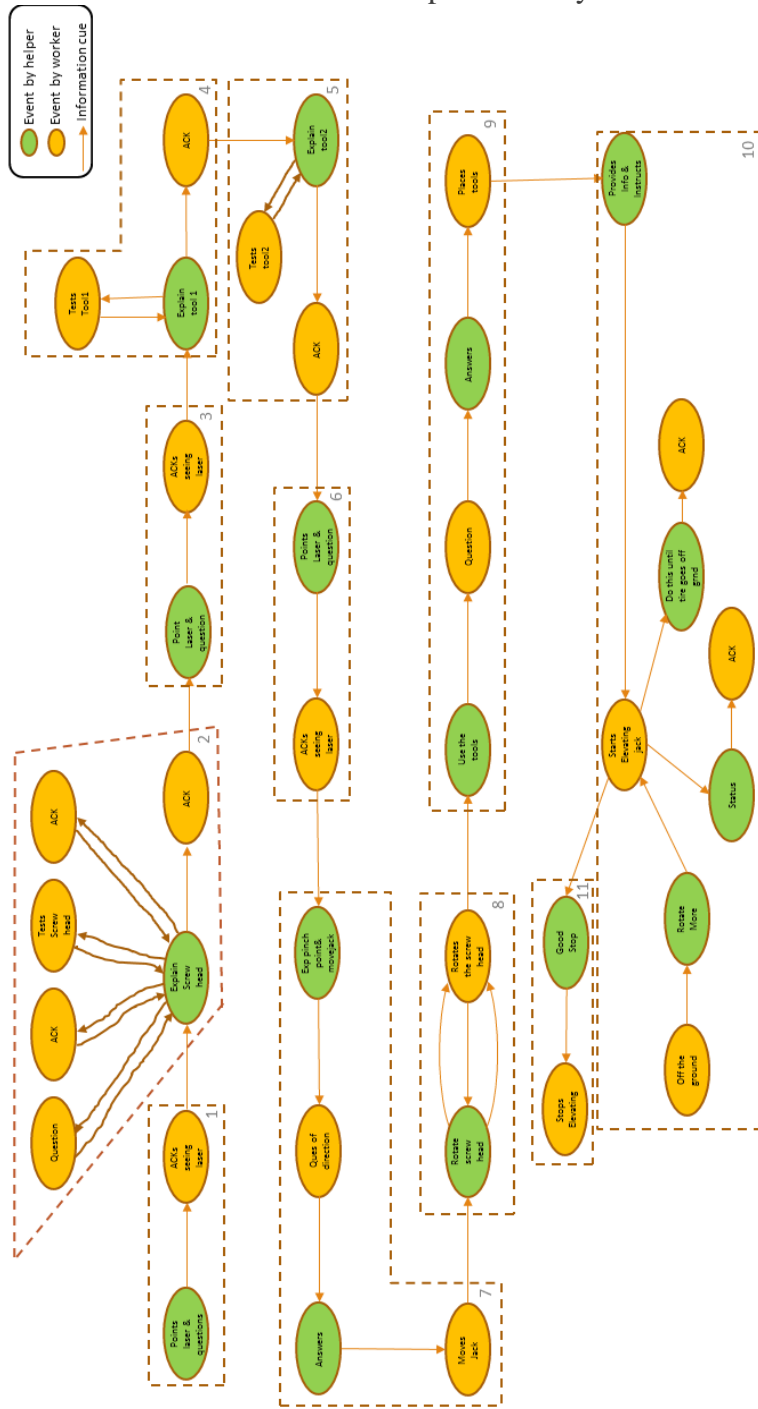

**Figure 28: Structural representation of a protocol**